\makeatletter
\newcommand{\dontusepackage}[2][]{%
  \@namedef{ver@#2.sty}{9999/12/31}%
  \@namedef{opt@#2.sty}{#1}}
\makeatother
\dontusepackage{subfigure}

\documentclass[]{article}

\usepackage{lmodern}
\usepackage{amssymb,amsmath}
\usepackage{ifxetex,ifluatex}
\usepackage[usenames,dvipsnames]{color}
\usepackage{fixltx2e} % provides \textsubscript
\ifnum 0\ifxetex 1\fi\ifluatex 1\fi=0 % if pdftex
  \usepackage[T1]{fontenc}
  \usepackage[utf8]{inputenc}
\else % if luatex or xelatex
  \ifxetex
    \usepackage{mathspec}
    \usepackage{xltxtra,xunicode}
  \else
    \usepackage{fontspec}
  \fi
  \defaultfontfeatures{Mapping=tex-text,Scale=MatchLowercase}
  
\fi
\IfFileExists{upquote.sty}{\usepackage{upquote}}{}
\IfFileExists{microtype.sty}{%
\usepackage{microtype}
\UseMicrotypeSet[protrusion]{basicmath} % disable protrusion for tt fonts
}{}
\usepackage[margin=1.0in,bottom=1.5in]{geometry}
\usepackage[square,numbers,sort&compress]{natbib}
\usepackage{listings}
\usepackage{graphicx}
\makeatletter
\def\maxwidth{\ifdim\Gin@nat@width>\linewidth\linewidth\else\Gin@nat@width\fi}
\def\maxheight{\ifdim\Gin@nat@height>\textheight\textheight\else\Gin@nat@height\fi}
\makeatother
\setkeys{Gin}{width=\maxwidth,height=\maxheight,keepaspectratio}
\usepackage{caption}
\usepackage{float}
\usepackage{subfig}
\ifxetex
  \usepackage[setpagesize=false, % page size defined by xetex
              unicode=false, % unicode breaks when used with xetex
              xetex]{hyperref}
\else
  \usepackage[unicode=true]{hyperref}
\fi
\hypersetup{breaklinks=true,
            bookmarks=true,
            pdfauthor={},
            pdftitle={Transfer learning in large-scale ocean bottom seismic wavefield reconstruction},
            colorlinks=true,
            citecolor=black,
            urlcolor=blue,
            linkcolor=black,
            pdfborder={0 0 0}}
\usepackage[preprint]{neurips_2019}
\usepackage{url}
\usepackage{booktabs}
\usepackage{amsfonts}
\usepackage{nicefrac}
\usepackage{microtype}

\title{Transfer learning in large-scale ocean bottom seismic wavefield
reconstruction}
\author{Mi Zhang\\School of Earth and Atmospheric Sciences,\\Georgia Institute
of Technology\\State Key Laboratory of Petroleum Resources and
Prospecting,\\China University of Petroleum -
Beijing\\\texttt{mzhang488@gatech.edu} \And
Ali Siahkoohi\\School of Computational Science and Engineering,\\Georgia
Institute of Technology\\\texttt{alisk@gatech.edu} \And
Felix J. Herrmann\\School of Earth and Atmospheric Sciences,\\School of
Computational Science and Engineering,\\Georgia Institute of
Technology\\\texttt{felix.herrmann@gatech.edu}}
\date{}

\begin{document}
\maketitle
\begin{abstract}
Achieving desirable receiver sampling in ocean bottom acquisition is
often not possible because of cost considerations. Assuming adequate
source sampling is available, which is achievable by virtue of
reciprocity and the use of modern randomized (simultaneous-source)
marine acquisition technology, we are in a position to train
convolutional neural networks (CNNs) to bring the receiver sampling to
the same spatial grid as the dense source sampling. To accomplish this
task, we form training pairs consisting of densely sampled data and
artificially subsampled data using a reciprocity argument and the
assumption that the source-site sampling is dense. While this approach
has successfully been used on the recovery monochromatic frequency
slices, its application in practice calls for wavefield reconstruction
of time-domain data. Despite having the option to parallelize, the
overall costs of this approach can become prohibitive if we decide to
carry out the training and recovery independently for each frequency.
Because different frequency slices share information, we propose the use
the method of transfer training to make our approach computationally
more efficient by warm starting the training with CNN weights obtained
from a neighboring frequency slices. If the two neighboring frequency
slices share information, we would expect the training to improve and
converge faster. Our aim is to prove this principle by carrying a series
of carefully selected experiments on a relatively large-scale
five-dimensional data synthetic data volume associated with wide-azimuth
3D ocean bottom node acquisition. From these experiments, we observe
that by transfer training we are able t significantly speedup in the
training, specially at relatively higher frequencies where consecutive
frequency slices are more correlated.
\end{abstract}

\section{Introduction}\label{introduction}

In seismic exploration, the complex and variable marine environment
brings about a unique set set of challenges to data acquisition. Because
we can safely assume that sources are sampled densely, by relying on
existing work on randomized marine acquisition
\citep{kumar2015sss, cheng2015separation}, our acquisition productivity
is dominated by attainable levels of sparsity in the distribution of
Ocean Bottom Nodes or Cables (OBN, OBC) without sacrificing the overall
quality of long-offset multi-azimuth data. Compared to other acquisition
methods, OBNs offer the most flexibility to deliver on this promise but
this comes with the challenge that we need to control costs deploying
OBNs by sampling the receivers extremely sparsely (at least $10\times$
subsampled).

This large degree of subsampling challenges most existing wavefield
reconstruction techniques that do not, either explicitly as in matrix or
tensor completion
\citep{lowrank2014Curt, lowrankc2015Rajiv, lowrank2016Oscar} or
implicitly as in recent work by \citet{siahkoohi2019seismic}, leverage
correlations that exist in monochromatic frequency slices across the
full survey area. The reason of this lies in the fact many approaches
\citep{lowrank2011Oropeza} rely on working in small upto five
dimensional windows where long-range correlations that exist in seismic
data volumes are ignored limiting their wavefield reconstruction
performance for wide-azimuth data. By working with monochromatic data
from across the whole survey, wide-azimuth wavefield recovery is
feasible for high degrees of subsampling as recently demonstrated by
\citet{lowrankc2015Rajiv}, \citet{lowrank2016Oscar}, and later
\citet{lowrankw2019Yijun}. In this work, explicit use is made during the
recovery of redundancies within monochromatic data that manifests itself
by the fact seismic data can be approximated in low-rank factored form
when organized in permuted form by lumping together sources/receivers in
$x$ and $y$ directions rather than combining source $x$ and source $y$
and receiver $x$ and receiver $y$. Because fully sampled frequency
slices are never formed explicitly, this approach has successfully been
applied to industry-scale problems \citep{lowrankc2015Rajiv} for the
low- to mid-frequency ranges. More accurate wavefield reconstruction at
higher frequencies has recently been made possible
\citep{lowrankw2019Yijun} via a recursive technique that sweeps from low
to high frequencies and where factorizations of neighboring (often at
lower temporal frequency) frequency slices are used in the recovery of
the current frequency slice. This weighting scheme is successful when
neighboring frequency slices have information in common with the current
frequency slice and recurrent application of this principle has resulted
in improvements of wavefield recovery at high frequencies from severely
subsampled data.

While (weighted) factored matrix completion techniques have been mainly
responsible for full-azimuth wavefield reconstruction from severe
subsampling, the low-rank factored approach is somewhat limited because
it essentially relies on a shallow (one layer) encoder-decoder
(linear)neural network---i.e., the low-rank factors can be thought as
neural net encoders decoders. However, from recent successes in machine
learning we know that deep convolutional neural networks (CNNs) are
capable of capturing more intricate relationships in the data. Judged by
the early success of \citet{siahkoohi2019seismic}, we ague that
relationships among the different gathers are captured implicitly by
training a Generative Adversarial Network \citep[GAN,][]{Goodfellow2014}
on pairs of fully sampled and subsampled monochromatic single-receiver
frequency slices. Compared to the earlier mentioned matrix-completion
approach, the latter approach is fundamentally nonlinear during which
similarities that live within the data are encoded in the weights of
network during training.

While GAN based wavefield reconstruction
\citep{siahkoohi2018seismic, siahkoohi2019seismic} can lead to
high-quality reconstructions, its computational costs, and therefore
performance, can become an issue especially when we move to higher
frequencies. This problem is exacerbated by the fact that each frequency
slice is treated independently---i.e., we train and reconstruct each
frequency slice separately. We present a method that overcomes this
problem by exploiting frequency-to-frequency similarities, in addition
to spatial redundancies that live across the monochromatic survey as a
whole. As during wavefield recovery with weighted factorizations, we use
information from neighboring frequency slices to inform training of the
GANs for the different frequencies through transfer training
\citep{yosinski2014transferable, transfer2019Ali}. We base this choice
for transfer training on positive experiences we have had using this
technique in different areas of seismic data processing and modeling
\citep{transfer2019Ali}. In these scenarios, transfer learning
significantly improved the wavefield reconstruction quality while
reducing training costs, specially at relatively higher frequencies
where consecutive frequency slices are more correlated.

Our paper is organized as follows. First, we discuss how to use
source-receiver reciprocity to construct training and testing data.
Second, we briefly introduce Generative Adversarial Networks
\citep[GANs,][]{Goodfellow2014}. Next, we explain how to use transfer
learning to finetune CNNs that are trained on neighboring frequencies to
reduce training costs. Finally, we demonstrate the performance of the
proposed method compared to state-of-the-art methods on a large-scale 5D
synthetic dataset.

\section{Extracting training pairs from
data}\label{extracting-training-pairs-from-data}

In the ocean bottom acquisition geometry discussed in this work, the
sources are assumed to be fully-sampled and the receivers are severely
subsampled. For this reason for each recorded receiver in the field, the
corresponding single-receiver frequency slice is fully sampled. On the
other hand, all single-source frequency slices are subsampled because of
the sparse OBN sampling.

We train our network to reconstruct monochromatic seismic data by
feeding it pairs of artificially subsampled (with a different
subsampling mask for each iteration of the training) and fully sampled
single-receiver frequency slices. During testing, the trained CNN is
used to recover missing values in single-source frequency slices---i.e.,
information in missing receivers. While not used explicitly, we made in
this approach use of reciprocity during training because we worked with
receiver gathers with dense source sampling.

\section{Network architecture and
optimization}\label{network-architecture-and-optimization}

During training of a GAN, the CNN, $\mathcal{G}_{\theta}$, which
performs the wavefield reconstruction, is coupled with an additional
CNN, the discriminator, $\mathcal{D}_{\phi}$, that learns to distinguish
between fully-sampled frequency slices and the ones that have been
recovered by $\mathcal{G}_{\theta}$. To enforce the relationship between
each specific pair of subsampled and fully-sampled frequency slices, we
include an additional $\ell_1$-norm misfit term weighted by $\lambda$
\citep{pix2pix2016}. We use the following objective function for
training GANs with input-output pairs:
\begin{equation}
\begin{aligned}
&\ \min_{\theta} \mathop{\mathbb{E}}_{\mathbf{X}\sim p(\mathbf{X})} \left [ \left (1-\mathcal{D}_{\phi} \left (\mathcal{G}_{\theta} (\mathbf{M} \odot \mathbf{X}) \right) \right)^2 + \lambda \left \| \mathcal{G}_{\theta} (\mathbf{M} \odot \mathbf{X})-\mathbf{X} \right \|_1 \right ] ,\\
&\ \min_{\phi} \mathop{\mathbb{E}}_{\mathbf{X}\sim p(\mathbf{X})} \left [ \left( \mathcal{D}_{\phi} \left (\mathcal{G}_{\theta}(\mathbf{M} \odot \mathbf{X}) \right) \right)^2 \ + \left (1-\mathcal{D}_{\phi} \left (\mathbf{X} \right) \right)^2 \right ],
\end{aligned}
\label{adversarial-training}
\end{equation}
 where $\mathbf{M}$ is the training mask, $\odot$ element-wise
multiplication, and the expectations are approximated with the empirical
mean computed over $\mathbf{X}_i, \ i = 1,2, \ldots , N_R$---i.e.,
fully-sampled single-receiver frequency slices drawn from the
probability distributions $p (\mathbf{X})$. As proposed by
\citet{johnson2016perceptual}, we use a ResNet \citep{he2016deep} for
the generator $\mathcal{G}_{\theta}$ and we follow \citet{pix2pix2016}
for the discriminator $\mathcal{D}_{\phi}$ architecture. We set the
hyper-parameter $\lambda$ as 1000 to balance generator's tasks for
fooling the discriminator and mapping specific pairs
$(\mathbf{M} \odot \mathbf{X}_i,\, \mathbf{X}_i)$ to each other
\citep{pix2pix2016}. Solving the optimization
objective~\ref{adversarial-training} is typically based on Stochastic
Gradient Descent (SGD) or one of its variants
\citep{Goodfellow-et-al-2016, bottou2018optimization}.

\section{Transfer learning between correlated
frequencies}\label{transfer-learning-between-correlated-frequencies}

Transfer learning involves utilizing the knowledge a neural network has
gained during pretraining in order to perform another but related task
\citep{TransferL2010Pan, transferL2012Bengio, transfer2019Ali}. In the
proposed deep-learning-based wavefield reconstruction framework, we
finetune weights of the CNN trained to reconstruct a neighboring
frequency component to reconstruct the slices of the current frequency
component. In case neighboring frequency slices are similar, this may
speed up the training compared to training a CNN from scratch.

Since the performance of transfer learning depends on the similarity
between tasks \citep{TL2014Similarity, TL2019Similarity}, it is best to
perform correlation analysis before transfer learning. To make this
qualitative statement more quantitative, we calculate the smallest
principal angles between row (or column) subspaces of two frequency
slices \citep{lowrankw2019Yijun}. Interested readers can refer to
\citet{lowrank2018Eftekhari} for an extensive overview of the
calculation. Small angles indicate a high correlation between two
slices. According to this calculation, the smallest angle value of row
subspaces is $0.11$ radian between $9.33$ Hz and $9.66$ Hz, whereas it
is $0.08$ radian between $14.33$ Hz and $14.66$ Hz. Similarly the
smallest angle value of column subspaces is $0.17$ radian between $9.33$
Hz and $9.66$ Hz, whereas it is $0.10$ radian between $14.33$ Hz and
$14.66$ Hz. Notwithstanding the fact that these angles are obtained
based on a linear factorization of the data, these values partially
support the fact that the correlation between two adjacent frequencies
$14.33$ and $14.66$ Hz is slightly higher than that between two
non-adjacent frequencies $9.33$ and $14.66$ Hz. For this reason, we
expect to see transfer learning to perform slightly more efficiently
when applied to finetune weights of the CNN trained to reconstruct
$14.33$ frequency slices to reconstruct the $14.33$ frequency slices.

\section{Numerical Experiments}\label{numerical-experiments}

To explore the reconstruction ability of the proposed method, we apply
it on a 5D synthetic dataset simulated to a portion of BG Compass model
with highly sparse receivers ($90\%$ of receivers are randomly missing)
and compare it with the low-rank matrix completion methods
\citep{lowrankc2015Rajiv, lowrankw2019Yijun}. The geometry is composed
of a $172 \times 172$ periodic grid of sources and a $172 \times 172$
periodic grid of receivers, both with $25$ m spatial sampling interval
in both $x$ and $y$ directions. We perform 1D fast Fourier transform
(FFT) to transform the seismic data from the time domain to the
frequency domain and then extract monochromatic $9.33$, $9.66$, $14.33$,
and $14.66$ Hz frequency components to showcase our method. For
different frequency components, we construct the corresponding training
and test sets according to the previously mentioned permutation. Then we
pretrain a randomly initialized CNN on all extracted monochromatic
frequency slices. Next, we employ transfer learning and use the CNN
weights trained to reconstruct monochromatic seismic slices at $9.33$
and $14.33$ Hz as an initial guess to train CNNs to reconstruct $9.66$
and $14.66$ Hz data. As mentioned before, during training (and transfer
learning), we change the training mask at every epoch, hence, each
training pair is only used once during optimization. Therefore, the
performance of the CNN over testing dataset (or validation set) can be
accurately approximated using the training data set. For this reason, we
safely calculate the SNR over training pairs at during training as the
metric to assess the reconstruction capability of network on test data.

\begin{figure*}
\centering
\subfloat[\label{results966Hzloss}]{\includegraphics[width=0.800\hsize]{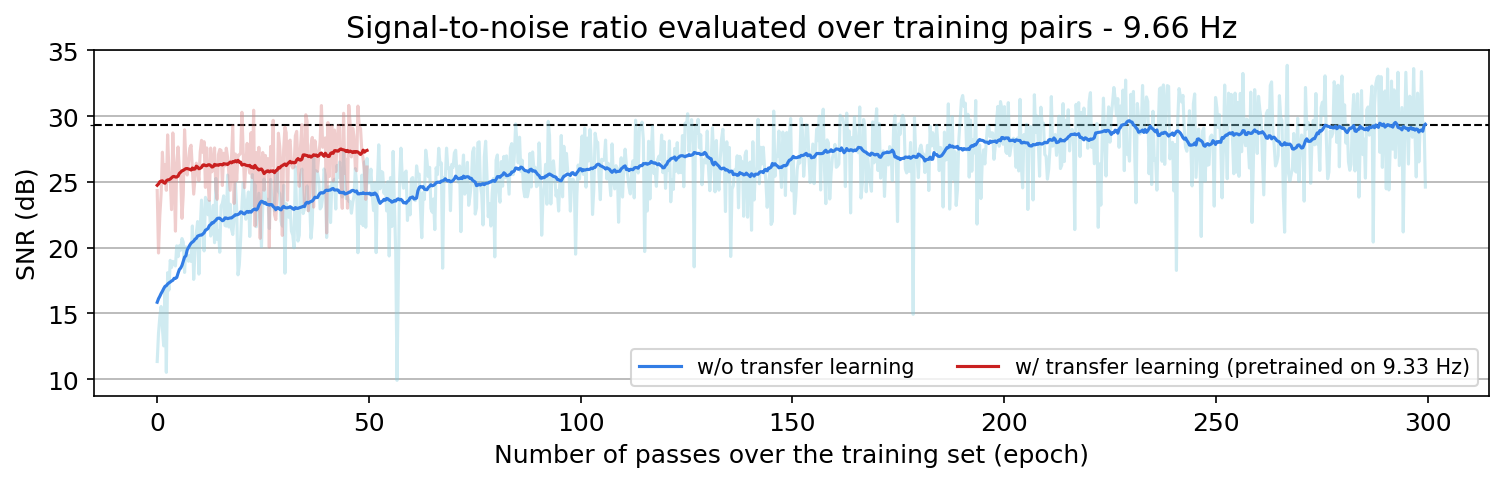}}
\\
\subfloat[\label{results1466Hzloss}]{\includegraphics[width=0.800\hsize]{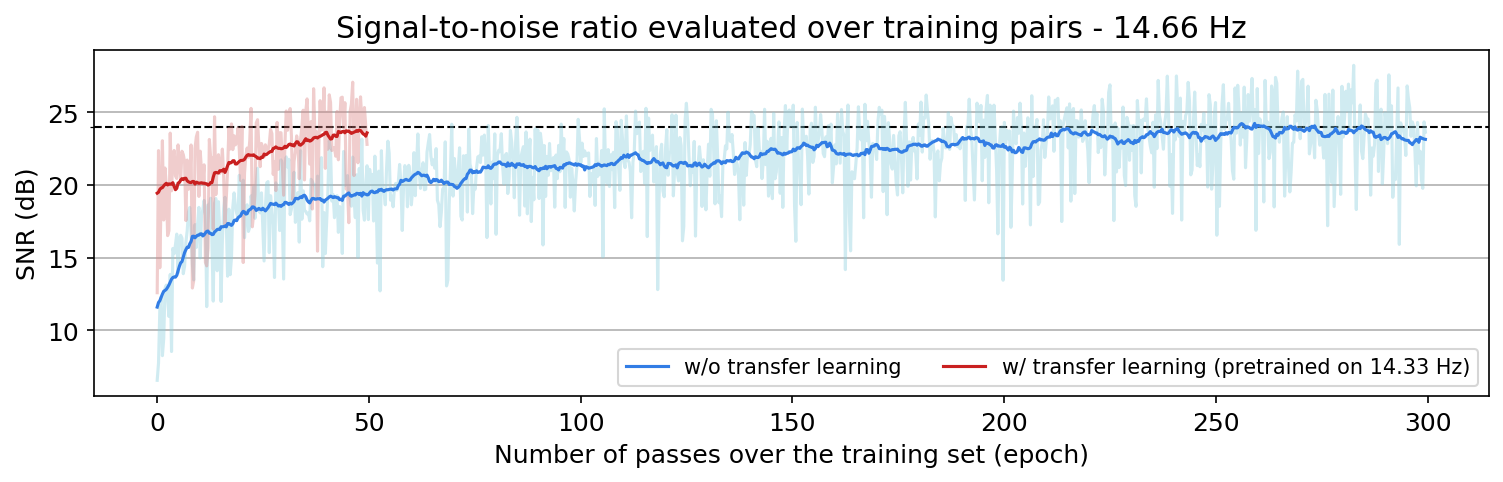}}
\caption{Transfer learning effectiveness when applied to neighboring and
non-neighboring (less similarity) frequency
slices.}\label{results1466Hzloss}
\end{figure*}

Figure~\ref{results966Hzloss} shows a comparison between $9.66$ Hz
frequency slice reconstruction SNRs, evaluated over training data during
training, using the original deep-learning based method
(light-blue)---i.e., training a randomly initialized, and result
obtained by transfer learning (light-red)---i.e., transferring a CNN
pretrained to reconstruct $9.33$ Hz slices frequency slices to
reconstruct 14.66 frequency slices. Similarly,
Figure~\ref{results1466Hzloss} shows the sampe comparison between for
$14.66$ Hz when we either train a randomly initialized CNN or apply
transfer learning using a CNN trained to reconstruct $14.33$ Hz data.
Dark colors indicate a running average over light curves to clarify the
overall trend. We can see that over $50$ epochs, the average SNR of
transfer learning of the CNN pretrained to reconstruct $14.33$ frequency
slices is always higher than that of the CNN directly trained from
scratch to reconstruct $14.66$frequency slices. We make a similar
observation in Figure~\ref{results966Hzloss} as well, except that
transfer learning needs more than $50$ epochs to obtain same
reconstruction SNR as the result without transfer learning. This
observation coincides with our expectation that transfer learning is
more effective when applied to more correlated tasks---i.e., when
neighboring frequency slices are more correlated. We also observed that
using transfer learning to reconstruct a neighboring frequency can
significantly speed up the the training, specially when consecutive
frequency slices are more correlated.

\begin{figure*}
\centering
\subfloat[\label{freq966Hz}]{\includegraphics[width=0.330\hsize]{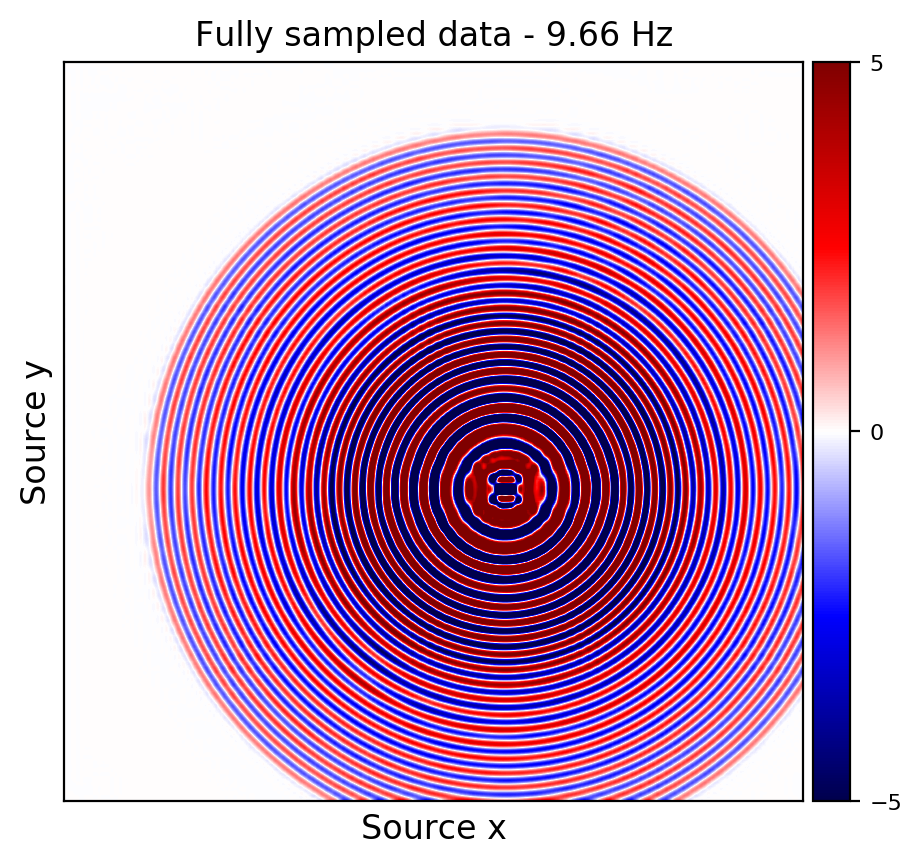}}
\subfloat[\label{freq966Hz-dl-error}]{\includegraphics[width=0.330\hsize]{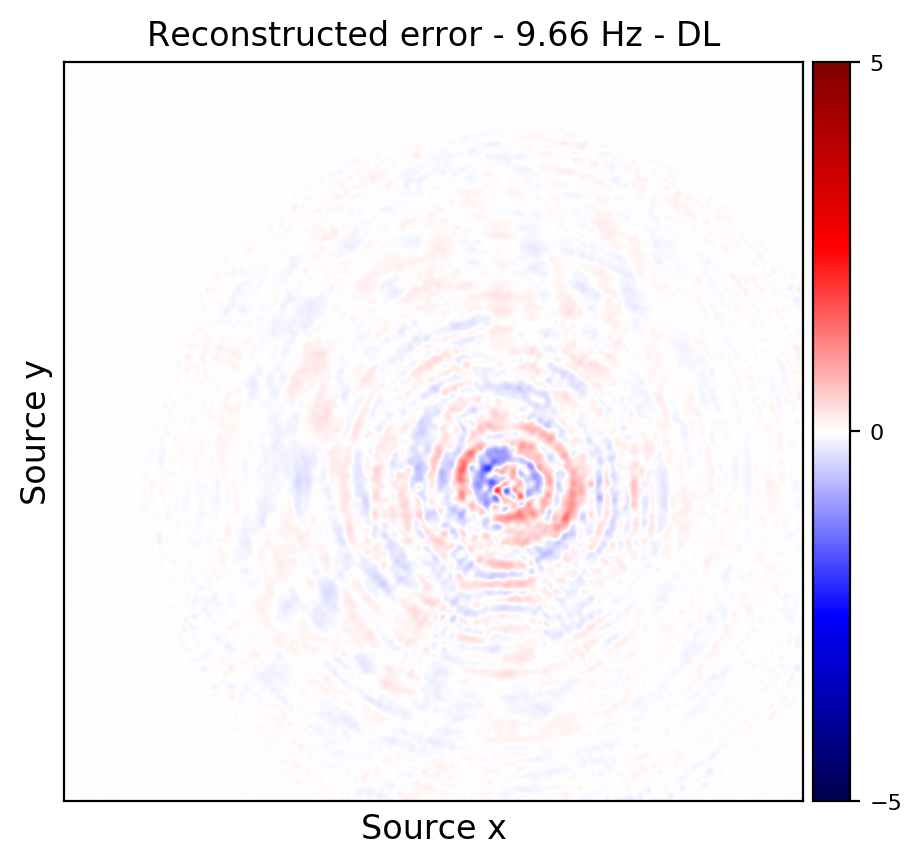}}
\subfloat[\label{freq966Hz-tl-error}]{\includegraphics[width=0.330\hsize]{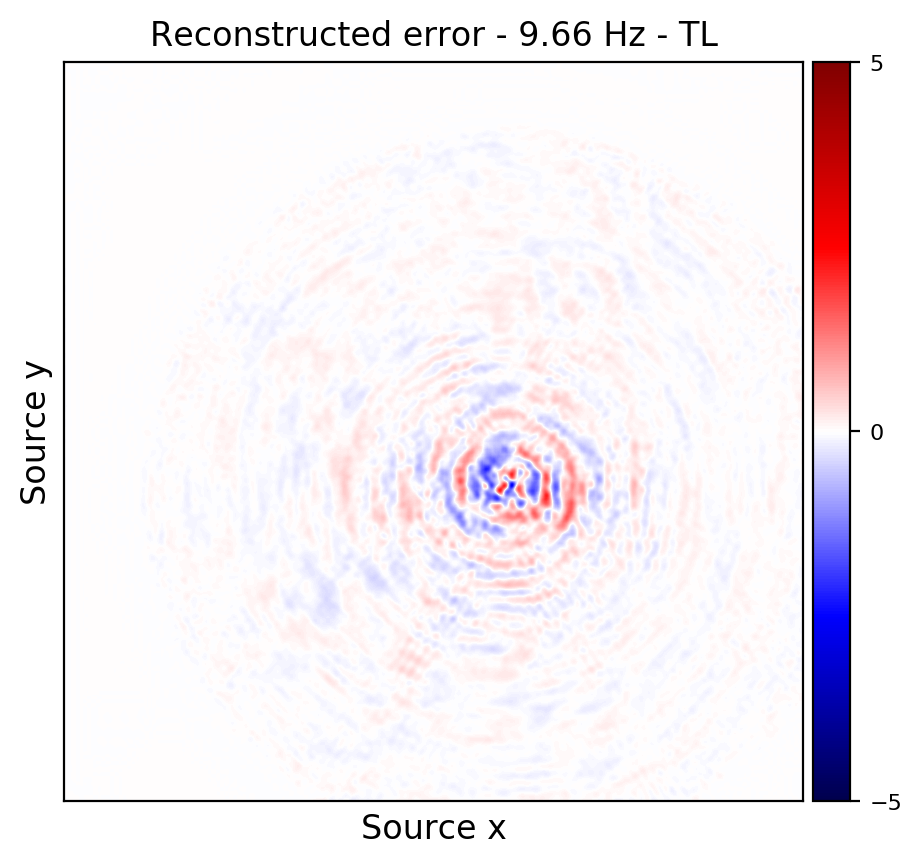}}
\\
\subfloat[\label{freq1466Hz}]{\includegraphics[width=0.330\hsize]{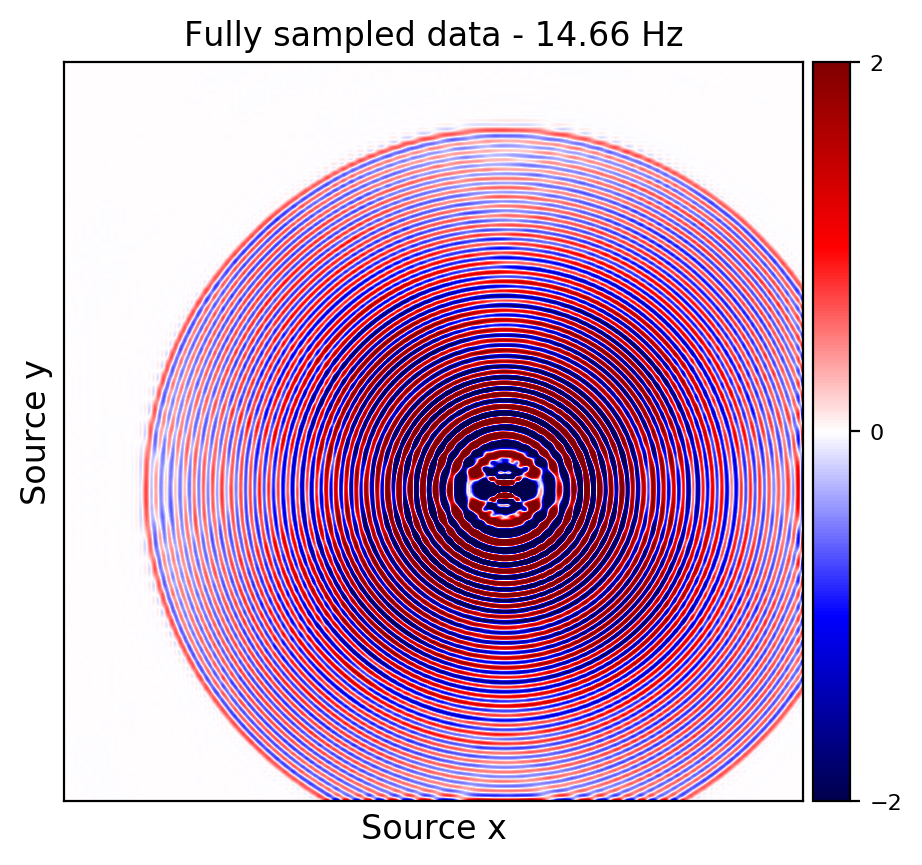}}
\subfloat[\label{freq1466Hz-dl-error}]{\includegraphics[width=0.330\hsize]{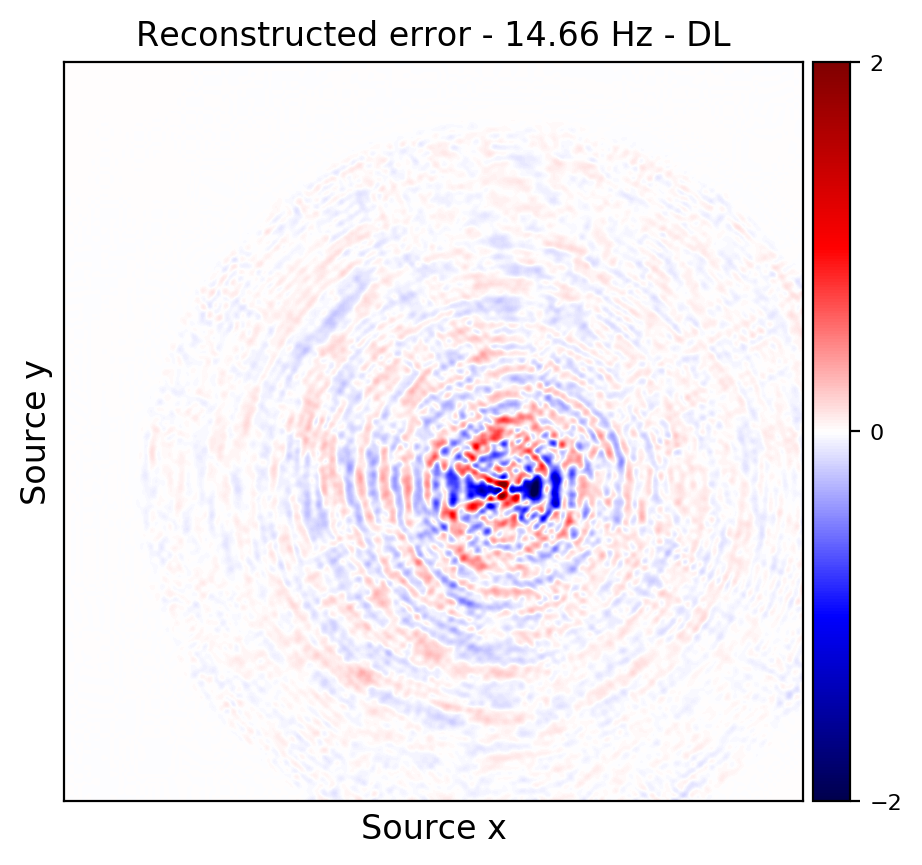}}
\subfloat[\label{freq1466Hz-dl-error}]{\includegraphics[width=0.330\hsize]{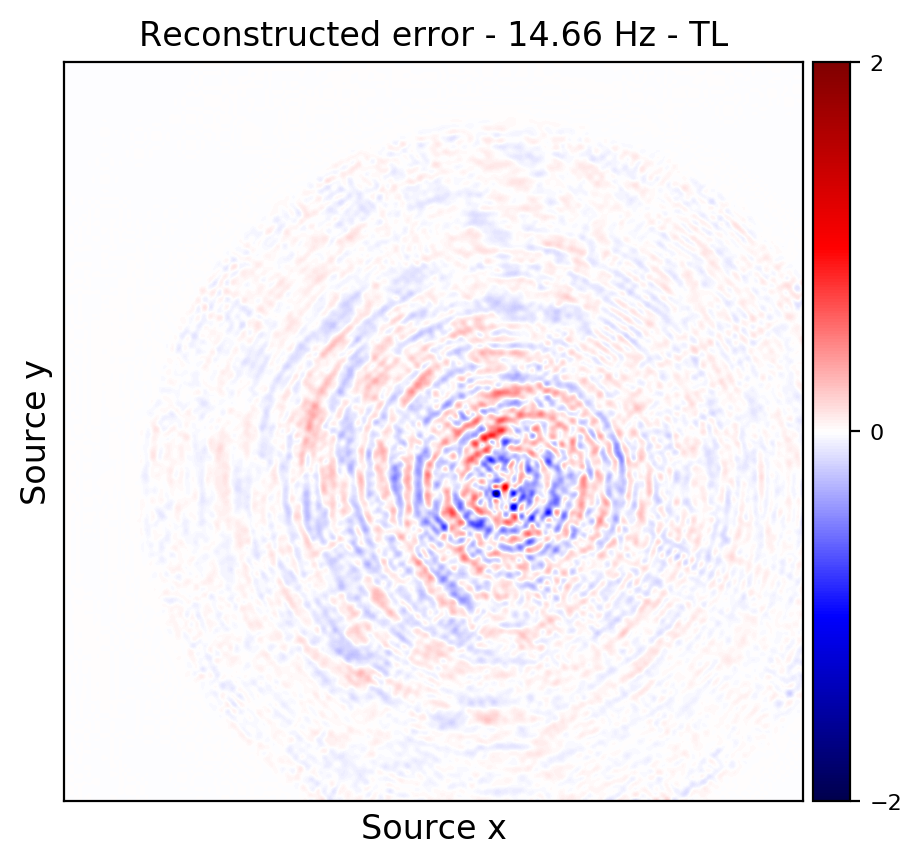}}
\caption{Reconstruction with and without transfer learning. (a-d) 9.66
and 14.66 Ground truth slices, respectively. (b, e) 9.66 and 14.66
recovery without transfer learning (c-f) 9.66 and 14.66 recovery with
transfer learning}\label{results966Hz}
\end{figure*}

Figures~\ref{freq966Hz} and~\ref{freq1466Hz} depict ground truth $9.66$
and $14.66$ Hz single-receiver frequency slices for a receiver that we
have assumed is missing in the observed data.
Figures~\ref{freq966Hz-dl-error} and~\ref{freq966Hz-tl-error} show the
reconstruction error obtained by training a randomly initialized CNN and
utilizing transfer learning to recover $9.66$ Hz data, respectively.
Similar figures for $14.66$ can be seen in
Figures~\ref{freq1466Hz-dl-error} and~\ref{freq1466Hz-dl-error}. As it
can be seen, transfer learning has been able to recover the slices with
similar quality, using much less computational cost. However, transfer
learning does a better job at recovering $14.66$ Hz data, which
coincides with our expectation given higher correlations among
consecutive frequency slices at higher frequencies.

\section{Conclusions}\label{conclusions}

In this work, we proposed to utilize transfer learning to improve the
training efficiency of our deep learning framework for seismic ocean
bottom wavefield reconstruction. Considering the similarities between
reconstruction tasks for frequency slices at neighboring frequencies, we
transfer the knowledge learned by the neural network for one frequency
to the other frequency. Our experiments on the 5D synthetic data
indicate that the knowledge transferred from adjacent frequencies is
reliable as long as the frequency slices share information. We found
that that is typically the case for higher frequencies that share more
information. We argue that this could be attributed to the fact that at
low frequencies the monochromatic slices are more orthogonal and
therefore share less information. Compared to our original deep-learning
based method, the proposed framework can speed up the training six fold
while improving the reconstruction performance.

\section{Related materials}\label{related-materials}

In order to facilitate the reproducibility of the results herein
discussed, a PyTorch \citep{NEURIPS2019_9015} implementation of this
work is made available on the
\href{https://github.com/slimgroup/Software.SEG2020/tree/master/zhang2020SEGtli}{GitHub}.

\bibliography{abstract}

\begin{thebibliography}{23}
\providecommand{\natexlab}[1]{#1}
\providecommand{\url}[1]{\texttt{#1}}
\expandafter\ifx\csname urlstyle\endcsname\relax
  \providecommand{\doi}[1]{doi: #1}\else
  \providecommand{\doi}{doi: \begingroup \urlstyle{rm}\Url}\fi

\bibitem[Kumar et~al.(2015{\natexlab{a}})Kumar, Wason, and
  Herrmann]{kumar2015sss}
Rajiv Kumar, Haneet Wason, and Felix~J. Herrmann.
\newblock Source separation for simultaneous towed-streamer marine acquisition
  {\textendash}- a compressed sensing approach.
\newblock \emph{GEOPHYSICS}, 80\penalty0 (6):\penalty0 WD73--WD88, 11
  2015{\natexlab{a}}.
\newblock \doi{10.1190/geo2015-0108.1}.
\newblock URL
  \url{https://slim.gatech.edu/Publications/Public/Journals/Geophysics/2015/kumar2015sss/kumar2015sss_revised.pdf}.

\bibitem[Cheng and Sacchi(2015)]{cheng2015separation}
Jinkun Cheng and Mauricio~D Sacchi.
\newblock Separation and reconstruction of simultaneous source data via
  iterative rank reduction.
\newblock \emph{GEOPHYSICS}, 80\penalty0 (4):\penalty0 V57--V66, 2015.

\bibitem[Silva and Herrmann(2014)]{lowrank2014Curt}
Curt~Da Silva and Felix~J. Herrmann.
\newblock Low-rank promoting transformations and tensor interpolation -
  applications to seismic data denoising.
\newblock In \emph{EAGE Annual Conference Proceedings}, 06 2014.
\newblock URL
  \url{https://slim.gatech.edu/Publications/Public/Conferences/EAGE/2014/dasilva2014EAGEhtucknoisy/dasilva2014EAGEhtucknoisy.pdf}.

\bibitem[Kumar et~al.(2015{\natexlab{b}})Kumar, Silva, Akalin, Aravkin,
  Mansour, Recht, and Herrmann]{lowrankc2015Rajiv}
Rajiv Kumar, Curt~Da Silva, Okan Akalin, Aleksandr~Y. Aravkin, Hassan Mansour,
  Benjamin Recht, and Felix~J. Herrmann.
\newblock Efficient matrix completion for seismic data reconstruction.
\newblock \emph{GEOPHYSICS}, 80\penalty0 (5):\penalty0 V97--V114,
  2015{\natexlab{b}}.
\newblock \doi{10.1190/geo2014-0369.1}.
\newblock URL \url{https://doi.org/10.1190/geo2014-0369.1}.

\bibitem[{López} et~al.(2016){López}, {Kumar}, {Yilmaz}, and
  {Herrmann}]{lowrank2016Oscar}
O.~{López}, R.~{Kumar}, {O}. {Yilmaz}, and F.~J. {Herrmann}.
\newblock Off-the-grid low-rank matrix recovery and seismic data
  reconstruction.
\newblock \emph{IEEE Journal of Selected Topics in Signal Processing},
  10\penalty0 (4):\penalty0 658--671, 2016.

\bibitem[Siahkoohi et~al.(2019{\natexlab{a}})Siahkoohi, Kumar, and
  Herrmann]{siahkoohi2019seismic}
Ali Siahkoohi, Rajiv Kumar, and Felix~J. Herrmann.
\newblock Deep-learning based ocean bottom seismic wavefield recovery.
\newblock \emph{SEG Technical Program Expanded Abstracts 2018}, Sep
  2019{\natexlab{a}}.
\newblock \doi{10.1190/segam2019-3216632.1}.
\newblock URL
  \url{http://www.earthdoc.org/publication/publicationdetails/?publication=92782}.

\bibitem[Oropeza and Sacchi(2011)]{lowrank2011Oropeza}
Vicente Oropeza and Mauricio Sacchi.
\newblock Simultaneous seismic data denoising and reconstruction via
  multichannel singular spectrum analysis.
\newblock \emph{GEOPHYSICS}, 76\penalty0 (3):\penalty0 V25--V32, 2011.
\newblock \doi{10.1190/1.3552706}.

\bibitem[{Zhang} et~al.(2019){Zhang}, Sharan, and Herrmann]{lowrankw2019Yijun}
Yijun {Zhang}, Shashin Sharan, and Felix~J. Herrmann.
\newblock High-frequency wavefield recovery with weighted matrix
  factorizations.
\newblock \emph{SEG Technical Program Expanded Abstracts 2019}, pages
  3959--3963, 2019.
\newblock \doi{10.1190/segam2019-3215103.1}.
\newblock URL
  \url{https://library.seg.org/doi/abs/10.1190/segam2019-3215103.1}.

\bibitem[Goodfellow et~al.(2014)Goodfellow, Pouget-Abadie, Mirza, Xu,
  Warde-Farley, Ozair, Courville, and Bengio]{Goodfellow2014}
Ian Goodfellow, Jean Pouget-Abadie, Mehdi Mirza, Bing Xu, David Warde-Farley,
  Sherjil Ozair, Aaron Courville, and Yoshua Bengio.
\newblock {G}enerative {A}dversarial {N}ets.
\newblock In \emph{Proceedings of the 27th International Conference on Neural
  Information Processing Systems}, NIPS'14, pages 2672--2680, 2014.
\newblock URL
  \url{http://papers.nips.cc/paper/5423-generative-adversarial-nets.pdf}.

\bibitem[Siahkoohi et~al.(2018)Siahkoohi, Kumar, and
  Herrmann]{siahkoohi2018seismic}
Ali Siahkoohi, Rajiv Kumar, and Felix~J. Herrmann.
\newblock {S}eismic {D}ata {R}econstruction with {G}enerative {A}dversarial
  {N}etworks.
\newblock \emph{80th EAGE Conference and Exhibition 2018}, Nov 2018.
\newblock \doi{10.3997/2214-4609.201801393}.
\newblock URL
  \url{http://www.earthdoc.org/publication/publicationdetails/?publication=92782}.

\bibitem[Yosinski et~al.(2014)Yosinski, Clune, Bengio, and
  Lipson]{yosinski2014transferable}
Jason Yosinski, Jeff Clune, Yoshua Bengio, and Hod Lipson.
\newblock How transferable are features in deep neural networks?
\newblock In \emph{Proceedings of the 27th International Conference on Neural
  Information Processing Systems}, NIPS'14, pages 3320--3328, 2014.
\newblock URL \url{http://dl.acm.org/citation.cfm?id=2969033.2969197}.

\bibitem[Siahkoohi et~al.(2019{\natexlab{b}})Siahkoohi, Louboutin, and
  Herrmann]{transfer2019Ali}
Ali Siahkoohi, Mathias Louboutin, and Felix~J. Herrmann.
\newblock The importance of transfer learning in seismic modeling and imaging.
\newblock \emph{GEOPHYSICS}, 84\penalty0 (6):\penalty0 A47--A52,
  2019{\natexlab{b}}.
\newblock \doi{10.1190/geo2019-0056.1}.
\newblock URL \url{https://doi.org/10.1190/geo2019-0056.1}.

\bibitem[Isola et~al.(2017)Isola, Zhu, Zhou, and Efros]{pix2pix2016}
Phillip Isola, Jun-Yan Zhu, Tinghui Zhou, and Alexei~A. Efros.
\newblock {I}mage-to-{I}mage {T}ranslation with {C}onditional {A}dversarial
  {N}etworks.
\newblock In \emph{The IEEE Conference on Computer Vision and Pattern
  Recognition (CVPR)}, pages 5967--5976, July 2017.
\newblock \doi{10.1109/CVPR.2017.632}.
\newblock URL \url{https://ieeexplore.ieee.org/document/8100115}.

\bibitem[Johnson et~al.(2016)Johnson, Alahi, and
  Fei-Fei]{johnson2016perceptual}
Justin Johnson, Alexandre Alahi, and Li~Fei-Fei.
\newblock {P}erceptual {L}osses for {R}eal-{T}ime {S}tyle {T}ransfer and
  {S}uper-{R}esolution.
\newblock In Bastian Leibe, Jiri Matas, Nicu Sebe, and Max Welling, editors,
  \emph{Computer Vision -- European Conference on Computer Vision (ECCV) 2016},
  pages 694--711, Cham, 2016. Springer International Publishing.
\newblock ISBN 978-3-319-46475-6.
\newblock \doi{10.1007/978-3-319-46475-6_43}.
\newblock URL
  \url{https://link.springer.com/chapter/10.1007%2F978-3-319-46475-6_43}.

\bibitem[He et~al.(2016)He, Zhang, Ren, and Sun]{he2016deep}
Kaiming He, Xiangyu Zhang, Shaoqing Ren, and Jian Sun.
\newblock {D}eep {R}esidual {L}earning for {I}mage {R}ecognition.
\newblock In \emph{The IEEE Conference on Computer Vision and Pattern
  Recognition (CVPR)}, pages 770--778, June 2016.
\newblock \doi{10.1109/CVPR.2016.90}.
\newblock URL \url{https://ieeexplore.ieee.org/document/7780459}.

\bibitem[Goodfellow et~al.(2016)Goodfellow, Bengio, and
  Courville]{Goodfellow-et-al-2016}
Ian Goodfellow, Yoshua Bengio, and Aaron Courville.
\newblock \emph{Deep Learning}.
\newblock MIT Press, 2016.
\newblock \url{http://www.deeplearningbook.org}.

\bibitem[Bottou et~al.(2018)Bottou, Curtis, and
  Nocedal]{bottou2018optimization}
L{\'e}on Bottou, Frank~E Curtis, and Jorge Nocedal.
\newblock Optimization methods for large-scale machine learning.
\newblock \emph{SIAM Review}, 60\penalty0 (2):\penalty0 223--311, 2018.

\bibitem[{Pan} and {Yang}(2010)]{TransferL2010Pan}
S.~J. {Pan} and Q.~{Yang}.
\newblock A survey on transfer learning.
\newblock \emph{IEEE Transactions on Knowledge and Data Engineering},
  22\penalty0 (10):\penalty0 1345--1359, Oct 2010.
\newblock ISSN 2326-3865.
\newblock \doi{10.1109/TKDE.2009.191}.

\bibitem[Bengio(2012)]{transferL2012Bengio}
Yoshua Bengio.
\newblock Deep learning of representations for unsupervised and transfer
  learning.
\newblock In Isabelle Guyon, Gideon Dror, Vincent Lemaire, Graham Taylor, and
  Daniel Silver, editors, \emph{Proceedings of ICML Workshop on Unsupervised
  and Transfer Learning}, volume~27 of \emph{Proceedings of Machine Learning
  Research}, pages 17--36, Bellevue, Washington, USA, 02 Jul 2012. PMLR.
\newblock URL \url{http://proceedings.mlr.press/v27/bengio12a.html}.

\bibitem[Ammar et~al.(2014)Ammar, Eaton, Taylor, Mocanu, Driessens, Weiss, and
  Tuyls]{TL2014Similarity}
Haitham~Bou Ammar, Eric Eaton, Matthew Taylor, Decebal~Constantin Mocanu, Kurt
  Driessens, Gerhard Weiss, and Karl Tuyls.
\newblock An automated measure of mdp similarity for transfer in reinforcement
  learning.
\newblock 2014.
\newblock URL
  \url{https://www.aaai.org/ocs/index.php/WS/AAAIW14/paper/view/8824}.

\bibitem[Dwivedi and Roig(2019)]{TL2019Similarity}
Kshitij Dwivedi and Gemma Roig.
\newblock Representation similarity analysis for efficient task taxonomy \&
  transfer learning.
\newblock pages 12379--12388, 06 2019.
\newblock \doi{10.1109/CVPR.2019.01267}.

\bibitem[{Eftekhari} et~al.(2018){Eftekhari}, {Yang}, and
  {Wakin}]{lowrank2018Eftekhari}
A.~{Eftekhari}, D.~{Yang}, and M.~B. {Wakin}.
\newblock Weighted matrix completion and recovery with prior subspace
  information.
\newblock \emph{IEEE Transactions on Information Theory}, 64\penalty0
  (6):\penalty0 4044--4071, 2018.

\bibitem[Paszke et~al.(2019)Paszke, Gross, Massa, Lerer, Bradbury, Chanan,
  Killeen, Lin, Gimelshein, Antiga, Desmaison, Kopf, Yang, DeVito, Raison,
  Tejani, Chilamkurthy, Steiner, Fang, Bai, and Chintala]{NEURIPS2019_9015}
Adam Paszke, Sam Gross, Francisco Massa, Adam Lerer, James Bradbury, Gregory
  Chanan, Trevor Killeen, Zeming Lin, Natalia Gimelshein, Luca Antiga, Alban
  Desmaison, Andreas Kopf, Edward Yang, Zachary DeVito, Martin Raison, Alykhan
  Tejani, Sasank Chilamkurthy, Benoit Steiner, Lu~Fang, Junjie Bai, and Soumith
  Chintala.
\newblock {PyTorch: An Imperative Style, High-Performance Deep Learning
  Library}.
\newblock In \emph{Advances in Neural Information Processing Systems 32}, pages
  8024--8035. 2019.
\newblock URL
  \url{http://papers.neurips.cc/paper/9015-pytorch-an-imperative-style-high-performance-deep-learning-library.pdf}.

\end{thebibliography}

\end{document}